\documentclass{epl}
\title{Forced motion of a probe particle near the colloidal glass transition}
\shorttitle{Forced motion of a probe particle ...}

\author{P. Habdas \and D. Schaar \and Andrew C. Levitt \and Eric R. Weeks}
\institute{Physics Department, Emory University, Atlanta, GA 30322}


\begin{document}

\maketitle

\begin{abstract}
We use confocal microscopy to study the motion of a magnetic
bead in a dense colloidal suspension, near the colloidal glass
transition volume fraction $\phi_g$.  For dense liquid-like
samples near $\phi_g$, below a threshold force
the magnetic bead exhibits only localized caged motion.
Above this force, the bead is pulled
with a fluctuating velocity.  The relationship between force
and velocity becomes increasingly nonlinear as $\phi_g$
is approached.  The threshold force and nonlinear drag force
vary strongly with the volume fraction, while the velocity
fluctuations do not change near the transition.
\end{abstract}

\section{Introduction}

Near the glass transition, molecular motion is greatly slowed,
yet the motion of molecules is difficult to directly observe and
the character of the molecular slowing is hard to determine
\cite{nagel1996,angel1995,stillinger1995,angell2000}.
Thus, colloidal suspensions of micrometer-sized
spheres are a useful model system for investigating
the nature of the glass transition, as they
can be directly observed using optical microscopy
\cite{kegel2000,weeks2000,weekscage2002,bartsch1998,marcus1999}.
As the volume fraction of a colloidal suspension is increased,
the motion of the particles becomes restricted due to the
confining effect of the other particles:  each particle is
temporarily trapped in a ``cage'' formed by its neighbors
\cite{megen1998,weekscage2002,saltzman2003,schweizer2003}.
For liquid-like samples, these cages eventually deform
and allow particles to diffuse through the sample
\cite{megen1998,weekscage2002}.  Above a critical volume
fraction ($\phi_g \sim 0.58$ for hard sphere colloids), the
particles' diffusion constant becomes zero and the macroscopic
viscosity increases dramatically \cite{cheng2002,megen1998}.
A system at such state is considered a colloidal glass and
$\phi_g$ is identified as the colloidal glass transition
point.  To gain insight into the nature of the colloidal
glass transition, past work used optical microscopy
to study the motion of individual particles 
as a function of volume fraction
\cite{kegel2000,weeks2000,weekscage2002,bartsch1998,marcus1999}.
These experiments discovered that particle motion in
equilibrated samples is both spatially heterogeneous
and temporally intermittent.  However, only a few studies
\cite{strating1999,wagner2001} have focused on nonequilibrium,
driven motion.

In this Letter, we study experimentally the motion of a
magnetic bead pulled through a dense colloidal suspension.
Similar to previous work with magnetic probes, we are able
to probe the local rheological properties of the suspension
\cite{leibler1996,sackmann2001,bacri2003}, although unlike
those studies, our samples are heterogeneous on the scale
of the magnetic probe.  In our experiments, as the colloidal
glass transition is approached the motion of the magnetic bead
becomes complex.  Below a threshold force, the magnetic bead
exhibits only localized caged motion.  When the external force is
above this threshold, the magnetic bead moves significantly
faster, with a fluctuating velocity.  Closer to $\phi_g$, the
threshold force rises, and the velocity-force relationship
becomes increasingly nonlinear.  Our measurements reveal
dramatic changes in the drag force acting on the magnetic bead
as the glass transition is approached. 

\section{Experimental}
The colloidal suspensions are made of poly-(methylmethacrylate)
particles, sterically stabilized by a thin layer of
poly-12-hydroxystearic acid \cite{antl1986}.  The particles
have a radius $a=1.10$ $\mu$m, a polydispersity of $\sim$5\%,
and are dyed with rhodamine.  The particles are slightly
charged, shifting the phase transition boundaries from those of
hard spheres.  We observe the freezing transition at $\phi_{f}
\approx$ 0.38, the melting transition at $\phi_{m} \approx$
0.42, and the glass transition at $\phi_{g} \approx$ 0.58.
The glass transition
is characterized by a vanishing diffusion constant for
the particles' motion, measured by confocal microscopy.
The colloidal particles are suspended in a mixture of
cyclohexylbromide/{\it cis}- and {\it trans}- decalin which
nearly matches both the density and the index of refraction
of the colloidal particles.  We add superparamagnetic beads
with a radius of $a_{\rm MB}=2.25$ $\mu$m (M450, coated with
glycidyl ether reactive groups, Dynal), at a volume
fraction $\sim 10^{-5}$, so that the magnetic beads
are well separated.  We do not observe attraction or repulsion
between the colloidal particles and the magnetic beads,
in either dilute or concentrated samples.  Before beginning
experiments, we mix the sample with an air bubble which
breaks up any pre-existing crystalline regions, then we wait 20
minutes before taking data, to allow the air bubble to stop
moving.  The colloidal suspension is not 
observed to flow during the acquisition of the data.

With a confocal microscope \cite{dinsmore2001}, we rapidly
acquire images (1.88 image/s) of area 80 $\mu$m $\times$ 80
$\mu$m, containing several hundred particles.  The magnetic
beads are not fluorescent and thus appear black on the
background of dyed colloidal particles.  To exert a force on
these beads, a Neodymium magnet is mounted on a micrometer held
just above the sample.  The force is calibrated by determining
the velocity of the magnetic beads in glycerol for a given
magnet position, and inferring the drag force from Stokes' Law
($F = 6 \pi \eta a_{\rm MB} v$, where $F$ is the drag force,
$\eta$ is the viscosity, and $v$ is the observed velocity)
\cite{sackmann2001}.  The imperfect reproducibility of the
magnet position causes a 5$\%$ uncertainty of the force.
Additionally, the variability in magnetic bead composition
results in an uncertainty in the magnetic force on different
beads, which we measure to be less than 10$\%$.  Also, the
magnetic beads are not density matched, and their effective
weight is 0.1 pN.  This is nearly negligible compared to
the applied horizontal forces in our experiments, and so in
our data below we consider only the applied forces and the
measured velocities in the $x$ direction.  We observe very
little vertical motion of the magnetic beads, typically less
than 5\% of the horizontal motion.  This is less than our other
measurement errors, discussed below.  We study isolated magnetic
beads at least 35 $\mu$m and more typically $>$50 $\mu$m from
the sample chamber boundary and from other magnetic beads.
Our experiments are limited by crystallization as the average
velocity of the magnetic bead decreases significantly in
the crystallized regions.  Thus, we make sure that the
data are collected well before crystallization appears.
Repeated measurements are reproducible before the onset of
crystallization.

\begin{figure}
\twofigures[scale=0.6]{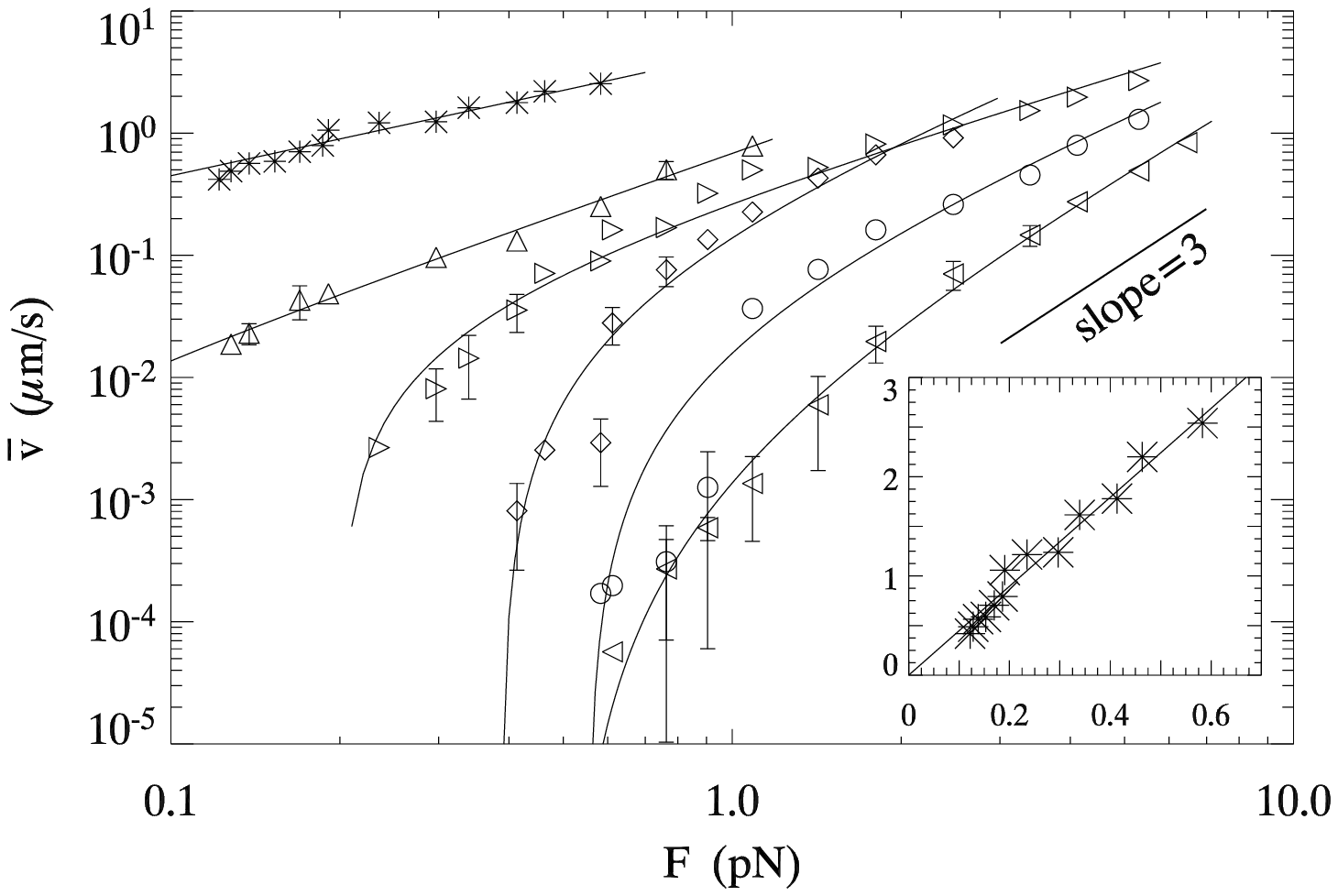}{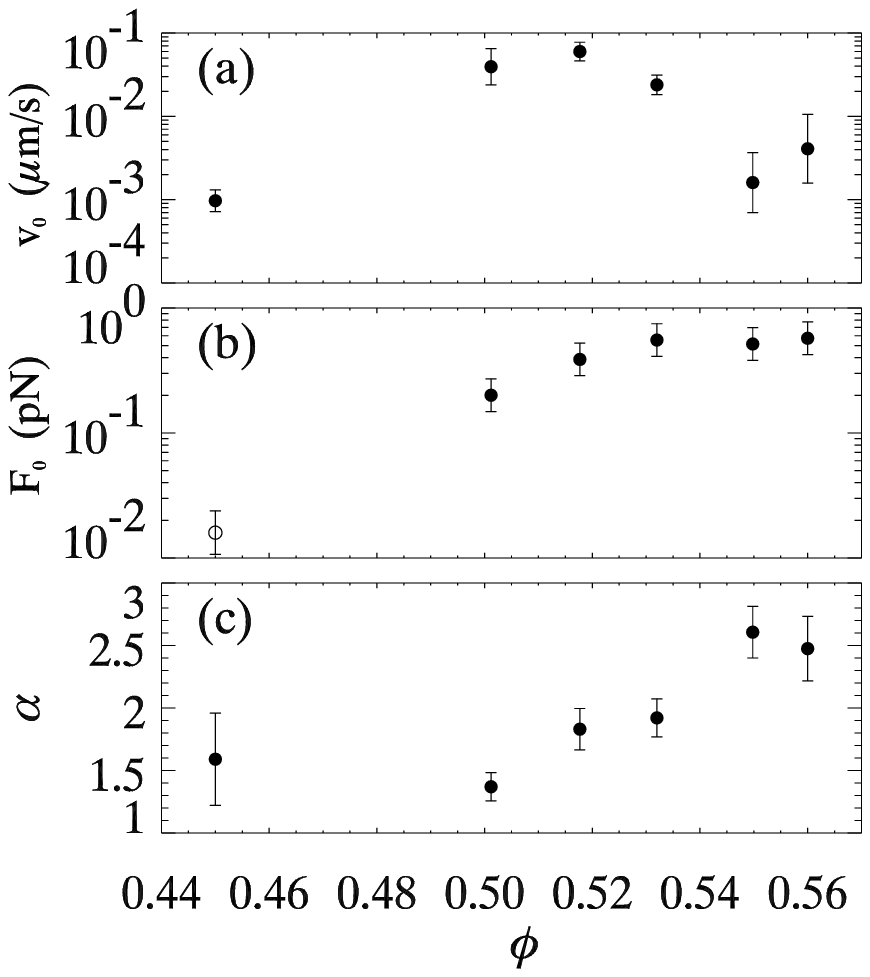}
\caption{Average velocity of the magnetic bead in the direction
of the motion as a function of force acting on the
magnetic bead for the volume fractions: 0.29 ($\ast$), 
0.45 ($\triangle$), 0.50 ($\triangleright$), 0.52
($\diamond$), 0.53 ($\circ$), 0.55 ($\triangleleft$).  
Solid lines are fits to the data obtained
using Eq.~(1).  Each data set is taken for a single
magnetic particle and thus the magnetic particle variability may
shift curves relative to each other but does not influence the
shape of a curve for a given volume fraction.  The velocity error
due to the uncertainty in the bead position is 5\%,
smaller than the symbol size.  
The slope=3 line is added for comparison.
Inset:  Data for $\phi=0.29$ on a linear/linear plot.
}
\label{fig1}
\caption{Fitting parameters as a function of volume fraction:
(a) $V_{0}$, (b) threshold force $F_{0}$, and (c) power
exponent $\alpha$.  The error bars are due to the uncertainties
in measuring the forces and velocities that the model is fit to.
The solid symbols for $F_0$ correspond to data sets for which stalled
behavior was directly observed below $F_0$ (all data sets $\phi
\geq 0.50$).
}
\label{fig2}
\end{figure}

\section{Results and discussion}
We study the motion of the magnetic beads for a range of forces.
Within each image, we locate the center of the magnetic bead
with an accuracy of at least 0.5 $\mu$m \cite{crocker1996}.
To show the behavior of the magnetic bead, we plot the average
magnetic bead velocity against the applied force on a log-log
plot in Fig.~\ref{fig1} for a range of volume fractions.
The average velocity of the magnetic bead for a given $F$
is given by $\bar{v} = \Delta x / \Delta t$ for the entire
course of the measurement; the error bars in Fig.~\ref{fig1}
are from the standard deviation of repeated measurements of
$\bar{v}$ for the same magnetic bead at the same $F$.  For a
dilute system ($\phi=0.29$, asterisks) the data fall on a line
with a slope of 1, indicating that the velocity is linearly
proportional to the applied force and Stokes' law applies
(Fig.~\ref{fig1} inset).  Due to the nonzero volume fraction,
the effective viscosity is modified from that of the solvent
($\eta_0 = 2.25$ mPa$\cdot$s) to the larger value of $\eta
= 6 \pm 1$ mPa$\cdot$s, in reasonable agreement with past
macroscopic measurements \cite{cheng2002}.

At higher volume fractions the velocity-force relationship
becomes more complex (Fig.~\ref{fig1}).  For data at large
forces, the average velocity seems to grow as a power law
with force, $\bar{v} \sim F^\alpha$, with exponent $\alpha$
between 1.5 and 3.  From left to right the curves increase in
volume fraction, and the slope at the highest forces for each
curve increases nearly monotonically, reflecting the approach
of the glass transition.  For all samples $\phi \geq
0.50$, we took data at extremely low forces, and found that
the magnetic bead did not move more than our resolution limit
within four hours; instead, it exhibits random caged motion.
This observation time is limited by the onset of crystallization
of the sample; in other words, the velocity is smaller than
$4 \cdot 10^{-5}$~$\mu$m/s.  Since the colloidal particles
move due to Brownian motion one would expect rearrangements to
``allow'' for the magnetic bead to move, even if this motion
would be primarily random with a small bias from the external
force.  However, within our observation time scale we cannot
see this behavior, and conclude that the velocity of such a
random walk is quite small.  For $\phi = 0.29$ and $0.45$, for
all forces checked, motion of the magnetic bead was observed.
(Forces below 0.1 pN are unobtainable, as this is the effective
weight of the magnetic bead.)

To determine how the velocity/force relationship changes as the
volume fraction is increased, we fit the data to the following equation:
\begin{eqnarray}
      \bar{v} =& v_{0} \cdot (\frac{F}{F_{0}} - 1)^\alpha, 
                  &F > F_0\\
            =& 0, &F \leq F_0\nonumber
\label{eq1}
\end{eqnarray}
where $F_{0}$ is a threshold force and $\alpha$ is the exponent
characterizing the growth of the velocity at large forces.
$v_0$ is the velocity the magnetic bead would have if the
applied force $F$ was equal to $2 F_0$.  This functional form
was also used to fit data from a two-dimensional simulation
similar to our experiments \cite{hastings2003}.  The solid lines
in Fig.~\ref{fig1} are the results of fitting the data with
Eq.~(1), with the exception of the $\phi=0.29$ data (asterisks)
where Stokes' Law applies.  While the model seems a good fit to
our data, our data are not strong enough to verify the model.
However, the model provides a useful way to quantify the main
features of the data.

As the colloidal glass transition is approached the system
becomes more viscous, and this is reflected in the changing
fit parameters from Eq.~(1) shown in Fig.~\ref{fig2}.
For low volume fractions, the three parameters
$v_0$, $F_0$, and $\alpha$ are all small, and Stokes' Law may
be the simplest explanation for the force/velocity data.
The fit parameters change as the volume fraction
$\phi\rightarrow\phi_g$.  $v_{0}$ initially grows with
increasing $\phi$ and then drops [Fig.~\ref{fig2}(a)].
The initial increase presumably reflects the change from
the $v_0 \rightarrow 0$ limit at low volume fractions.
The decrease as the glass transition is approached may be
due to the sharply increasing effective viscosity acting on
the magnetic bead. $v_0$ is also sensitive
to variations in the other fitting parameters,
as for large $F$, $v \approx (v_0 F_0^{-\alpha}) F^\alpha$.

The threshold force $F_0$ grows as $\phi$ increases, reaching
a plateau value for large $\phi$.  We expect that at $\phi_g$
the threshold force will be finite, both from our data,
and also as it seems more likely that $F_0$ diverges at
$\phi_{\rm RCP} \approx 0.64$, the random close-packing
volume fraction \cite{cheng2002}.  The value of the threshold
force near $\phi_g$, $F_0 \sim 1$ pN, is similar to that
predicted by the work of Schweizer and Saltzman {\it et
al.} \cite{saltzman2003,schweizer2003}.  They determined an
effective free energy characterizing caged particle motion,
and the derivative of this energy can be used to find an
effective force to break out of a cage, resulting in a value
of $\approx 50 k_B T / a \approx 0.2$ pN for $\phi = 0.6$
\cite{saltzman2003,schweizer2003}.  This prediction considers
the caged particle to be the same size as the surrounding
particles, and for our larger magnetic beads the stall force
would likely be larger, closer to our experimental values.
Thus, our measured threshold forces $F_0$ may be related to
the strength of colloidal cages; previously only the sizes of
these cages have been measured \cite{megen1998,weekscage2002}.

The exponent $\alpha$ changes dramatically as the glass
transition is approached, as seen in Fig.~\ref{fig2}(c),
where it rises to almost 3 near $\phi_g$.  Our results are in
contrast to Ref.~\cite{hastings2003}; while the model fits
their data as well, they found $\alpha = 1.5$ for glassy
systems.  The difference may be due to the much different
particle interactions: they studied unscreened point charges in
a 2D system, qualitatively different from our colloidal particles.  
In our experiments, it is unclear if $\alpha$ diverges at $\phi_g$
or continues to grow as $\phi_{\rm RCP}$ is approached.
Experiments at $\phi > \phi_g$ are difficult to interpret due to
the aging of the sample \cite{courtland2003}: for a given force
the magnetic bead response depends on the age of the sample.

\begin{figure}
\onefigure[scale=0.5]{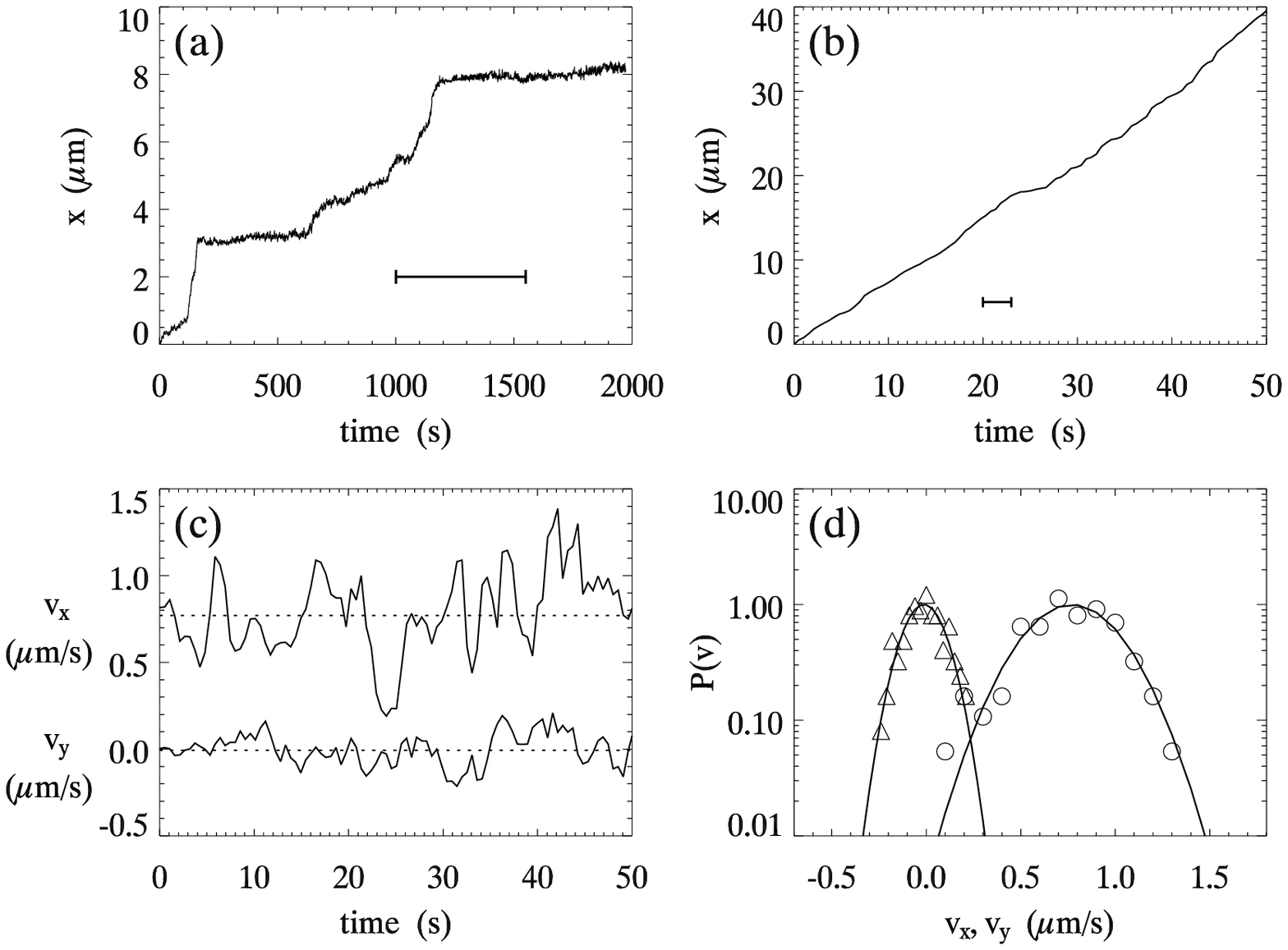}
\caption{(a) The position of the magnetic bead in the direction
of the motion as a function of time, $F=0.58$ pN, $\phi=0.52$,
$v=0.0041$ $\mu$m/s.  The scale bar corresponds to $a_{\rm MB}/v=550$s.
(b) The position of the magnetic bead in the direction
of the motion as a function of time, $F=6.46$ pN, $\phi=0.55$,
$v=0.80$ $\mu$m/s.  The scale bar corresponds to
$a_{\rm MB}/v=3$ s.
(c) The instantaneous velocity in the direction of motion
$v_{x}$ and perpendicular to the motion $v_{y}$ as a function
of time.  The velocities are calculated from the displacement
data using $\Delta t=3$ s.  The dashed lines denote the average
velocities $\bar{v_x}=0.80$ $\mu$m/s and $\bar{v_y}=0.01$
$\mu$m/s.  The data are the same as (b).
(d) The probability distribution of the instantaneous velocities
$v_{x}$ ($\circ$) and $v_{y}$ ($\triangle$).  The solid lines
denote a Gaussian fit with $\sigma_{vx}=0.24$~$\mu$m/s and 
$\sigma_{vy}=0.10$~$\mu$m/s.  
The data are the same as (b).}
\label{fig3}
\end{figure}

The motion of the magnetic bead is not smooth, as is seen
by a typical plot of its position as a function of time
[Fig.~\ref{fig3}(a,b)].  While the largest ``jumps'' seen in
Fig.~\ref{fig3}(a) are roughly the diameter of the surrounding
colloidal particles ($2a \sim 2$ $\mu$m), at other times the
magnetic bead takes much smaller steps.  When the magnetic
bead moves faster the fluctuations become less significant
and the motion of the magnetic bead is smoother, shown in
Fig.~\ref{fig3}(b).

We arbitrarily choose $\Delta t = a_{\rm MB}/\bar{v}$ as a time
scale and define the instantaneous velocity as $\vec{v}(t) =
[\vec{r}(t+\Delta t) - \vec{r}(t)] / \Delta t$.  This time scale
is indicated with the scale bars shown in Fig.~\ref{fig3}(a,b),
and the results that follow are not sensitive to this choice
(and in particular do not change using $a$ rather than $a_{\rm
MB}$).  The instantaneous velocity is shown in
Fig.~\ref{fig3}(c).  The magnetic bead velocity has much larger
fluctuations along the direction of motion [$v_x$, top trace
in Fig.~\ref{fig3}(c)], while fluctuations in the transverse
direction are smaller.  The distributions
of $v_{x}$ and $v_{y}$, shown on Fig.~\ref{fig3}(d), are
Gaussian (solid fit lines).  The fluctuations in $x$ and $y$ 
appear to be only weakly correlated.  The velocity fluctuations reflect
the fact that the magnetic bead is of similar size to the
colloidal particles: its motion is sensitive to colloidal
particle configurations.  Furthermore, dense colloidal suspensions
are spatially heterogeneous systems \cite{kegel2000,weeks2000} and 
some regions may be ``glassier'' and thus harder to move through.

\begin{figure}
\onefigure[scale=0.5]{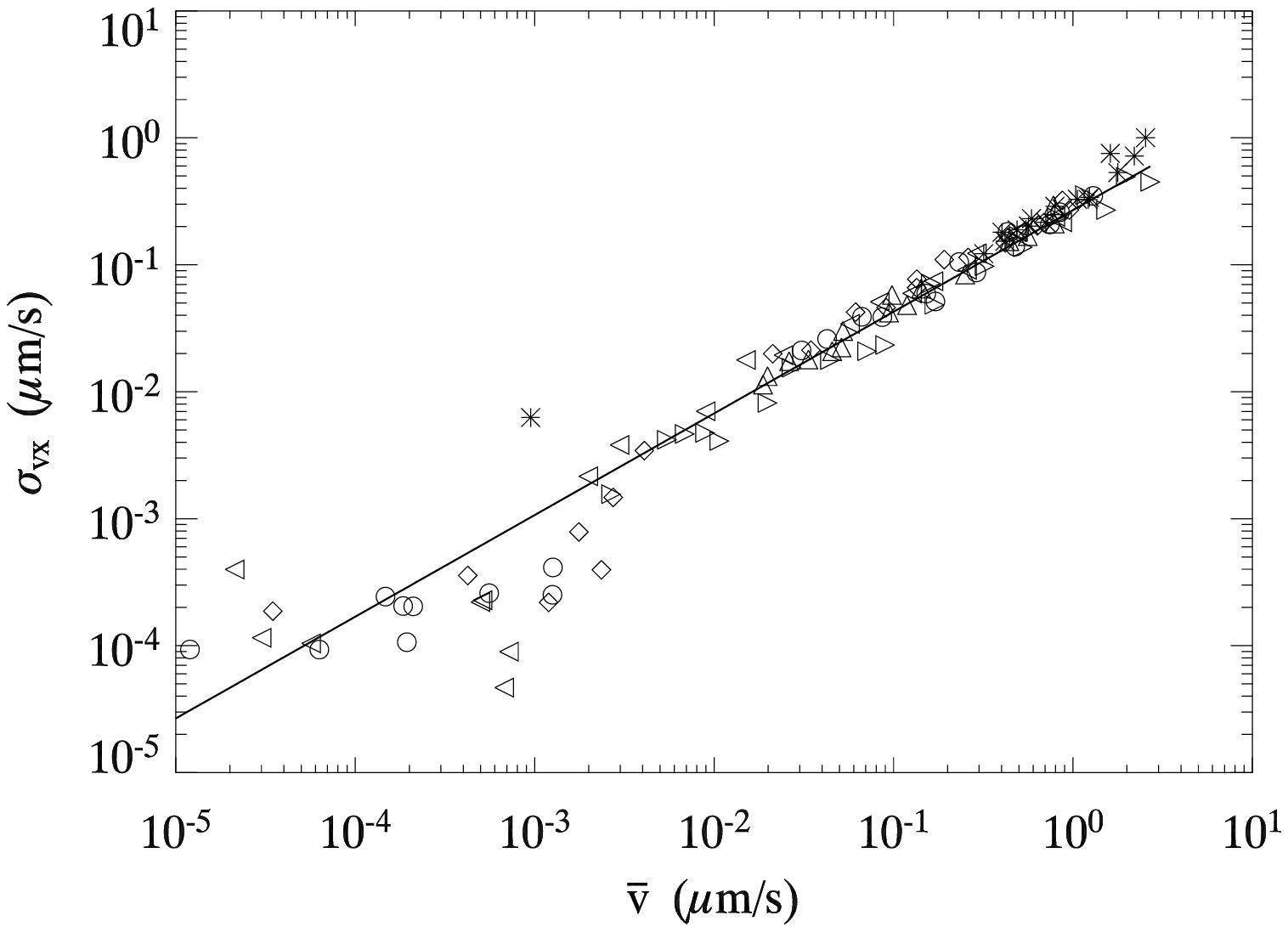}
\caption{The standard deviation of the instantaneous
distribution of $v_{x}$, ($\sigma_{vx}$) as a function
of $\bar{v}$.  The solid line is a fit to the data and has
a slope of 0.80.  The symbols represent the different volume
fractions which are the same as Fig.~\ref{fig1}.}
\label{fig4}
\end{figure}

As the glass transition is approached, the spatial heterogeneity
increases \cite{kegel2000,weeks2000}, and one might expect that
the velocity fluctuations would become more significant.
However, this is not the case, as is seen in Fig.~\ref{fig4} which
shows the standard deviation of the instantaneous velocity
$\sigma_{vx}$ plotted against the average velocity $\bar{v}$.
Different symbols indicate different volume fractions, showing
that it is not the volume fraction which determines the standard
deviation but the average velocity:  the fluctuations are larger
when the average velocity is larger.
Denser suspensions have slower velocities for the same force
(as seen in Fig.~\ref{fig1}) but also correspondingly smaller
velocity fluctuations.  In Fig.~\ref{fig4}, the data are more scattered at low
velocities and for any given sample the data scatter on both
sides of the fit line.  For different choices of $\Delta t$, we find that
$\sigma_{vx} \sim \bar{v}^{\beta}$ with the exponent $\beta = 0.8
\pm 0.1$.  The smaller the $\Delta t$ the wider distribution of $\sigma_{vx}$.
However, the general dependence of $\sigma_{vx}$ on $\bar v$ does not depend
on $\Delta t$.  We find that $\sigma_{vy} \approx 0.4 \sigma_{vx}$ and 
thus a similar trend holds for the transverse fluctuations.

The changing character of the motion as the glass transition
is approached can be understood by calculating the modified
Peclet number.  This is the ratio of two time scales,
$\tau_{\rm B}$ characterizing the unforced motion of the
colloidal particles, and $\tau_{\rm M}$ characterizing the
motion of the magnetic bead.  The Brownian time $\tau_{\rm
B}=a^2/2D_\infty$ is the time it takes for colloidal particles
to diffuse a distance equal to their radius $a$ and is based
on their asymptotic diffusion constant $D_\infty$.  $D_\infty$
becomes small near the glass transition and becomes zero for a
colloidal glass.  The magnetic bead time scale $\tau_{\rm M}
= a/\bar{v}$ is the time scale for the magnetic bead to move
the distance $a$ and is based on the magnetic bead's average
velocity $\bar{v}$.  The modified Peclet number is $Pe^* =
\tau_B / \tau_M$ and is larger than 1 for all symbols shown
in Fig.~\ref{fig1} \cite{footnote1}.  For the largest forces,
$Pe^* \approx 10000$.  Thus, the forced motion of the magnetic
bead is much more significant than the Brownian motion of the
surrounding colloidal particles: the magnetic bead pushes
these particles out of the way, plastically rearranging
the sample.  This results in a lowering of the effective
viscosity acting on the magnetic bead at higher velocities.
On the macroscopic scale, this may correspond to traditional
rheological measurements at high strains.  Past work found
for similar colloidal samples that the viscoelastic moduli
decreased as the maximum strain increased \cite{mason95}.

For extremely low forces, $Pe < 1$ and the velocity should
be linearly related to the force.  However, the velocities
predicted by linear response theory are well below what we
observe (and for low forces, below what we can measure).
This is not surprising, as the lowest forces we can apply are
fairly large; a nondimensional way to characterize this is the
ratio $a_{\rm MB} F/k_B T$  which is $\sim 50$ for $F$=0.1 pN.

Since $Pe^* > 1$ in our experiments, the Brownian motion should
be unimportant, similar to a granular system.  Moreover,
conjectures of the existence of a ``jamming transition''
speculate that the colloidal glass transition is similar
to jamming in granular media \cite{liu1998}, and so it is
interesting to compare our results with studies performed in
a granular media.  Albert {\it et al.} \cite{schiffer1999}
immersed a cylinder in granular particles and measured the
force exerted on the cylinder when it moved with constant speed
relative to the particles.  They found that the drag force on
the cylinder is independent of the velocity, quite different
from our behavior (Fig.~\ref{fig1}).  In our experiments,
the solvent viscosity may be important, causing the velocity
dependent drag force.

In conclusion, we find that near the glass transition the
motion of a microsphere through a colloidal suspension changes
dramatically.  A threshold force appears, below which the
motion of the sphere is localized, and this threshold force
increases as the glass transition is approached.  The existence
of the threshold force hints that the system may locally
``jam'' even when the colloidal suspension is globally still
a liquid.  Above the threshold force the velocity is related
to the force by a power law and the sphere moves with a
fluctuating velocity, locally deforming the sample.  At high
forces, the velocity is related to the force by a power law.
This can be inverted to obtain an effective drag force on the
sphere, growing weakly with velocity, $F \sim v^{1/\alpha}$
with $\alpha$ growing from 1 far from the glass transition to
nearly 3 close to the transition.  These results indicate that
the approach of the colloidal glass transition is signaled
by a growing nonlinearity of the drag force.  This may be
a new way to characterize the glass transition and should
serve as a useful test for some glass transition theories
\cite{saltzman2003,schweizer2003,falk,barrat2002}.

We thank R.~E.~Courtland, M.~B.~Hastings, S.~A.~Koehler,
D.~Nelson, K.~S.~Schweizer, T.~Squires, T.~A.~Witten, and S.~Wu for helpful
discussions.  We thank A.~Schofield for providing our colloidal
samples.  This work was supported by NASA (NAG3-2284).




\begin{thebibliography}{0}
\bibitem{footnote1} We term this the ``modified'' Peclet number
as we are using $D_\infty$ rather than $D_0$, the diffusion
constant characterizing particle motion in dilute samples.
\end{thebibliography}
\end{document}